\documentclass{elsart}%
\usepackage{amsfonts}
\usepackage{amsmath}
\usepackage{amssymb}
\usepackage{graphicx}%
\begin{document}
%
\begin{frontmatter}%
%

\title{Burr, Lévy, Tsallis}%
%

\author{F.Brouers(a,b), O.Sotolongo-Costa(b) K.Weron(c)}%
%

\address{(a) Department of Physics, University of Liège, 4000, Belgium}%
%

\address
{(b) Faculty of Physics, Chair of Complex Systems H.Poincaré, University of Havana, Cuba}%
%

\address
{(c) Institute of Physics, Wroclaw University of Technology, 50.370 Poland}%
%

\begin{abstract}
The purpose of this short paper dedicated to the 60th anniversary of
Prof.Constantin Tsallis is to show how the use of mathematical tools and
physical concepts introduced by Burr, L\.{e}vy and Tsallis open a new line of
analysis of the old problem of non-Debye decay and universality of relaxation.
We also show how a finite characteristic time scale can be expressed in terms
of a $q$-expectation using the concept of $q$- escort probability.The
comparison with the Weron et al. probabilistic theory of relaxation leads to a
better understanding of the stochastic properties underlying the Tsallis
entropy concept.
\end{abstract}%
%

%

\end{frontmatter}%

\section{Maximum entropy principle and probability distributions}

Most of the probability distributions used in natural, biological, social and
economic sciences can be formally derived by maximizing the entropy with
adequate constraints ($maxS$\ principle)\cite{Kap89}.

According to the $maxS$\ principle, given some partial information about a
random variable $i.e.$\ the knowledge of related macroscopic measurable
quantities (macroscopic observables), one should choose for it the probability
distribution that is consistent with that information but has otherwise a
maximum uncertainty.\ In usual thermodynamics, the temperature is a
macroscopic observable and the distribution functions are exponentials.

Quite generally, one maximizes the Shannon-Boltzmann (S-B) entropy:
\begin{equation}
S=-\int_{a}^{b}f(x)\ln f(x)dx
\end{equation}
subject to the conditions%

\begin{equation}
\int_{a}^{b}f(x)dx=1,\text{ \ }\int_{a}^{b}g_{i}(x)f(x)dx=<g_{i}(x)>,\text{
\ }i=1,2,....
\end{equation}
Both limits $a$ and $b$ may be finite or infinite.\ The functions $g_{i}(x)$
whose expectation value have been usually considered \cite{Kap89} as
constraints to build probability distributions are of the type
\begin{equation}
x,x^{2},x^{n},(x-<x>)^{2},\mid x\mid,\mid x-<x>\mid,\ln x,(\ln x)^{2},\ln(1\pm
x),\exp(-x)....
\end{equation}
The maximum entropy probability density function ($mepdf)$ depends on the
choice of the limits of integration $a$ and $b$ and the functions $g_{i}(x)$
whose expectation values are prescribed.

One constructs the Lagrangian
\[
L=-\int_{a}^{b}f(x)\ln(f(x)dx
\]

\begin{equation}
-\lambda_{0}(\int_{a}^{b}f(x)dx-1)-\sum_{i}\lambda_{i}(\int_{a}^{b}%
g_{i}(x)f(x)dx-<g_{i}(x)>)\
\end{equation}
and differentiating with respect to $f(x)$ the method of Lagrange multiliers
leads to :%

\begin{equation}
f(x)=C\exp[-\sum_{i}\lambda_{i}g_{i}(x)]
\end{equation}
The factor $C$ is a normalization constant and the Lagrange parameters
$\lambda_{i}$ are determined by using the constraints (3).When this cannot be
achieved simply, the parameters $\lambda_{i}$ are parameters defined by the
constraints. Most distributions derived from the constraints given in (3)
posses finite second moments and hence belong to the domain of attraction of
the normal distributions.\ Those which belong to the domain of attraction of
the L\.{e}vy (stable) distribution $i.e$. the Cauchy and the Pareto
distributions are obtained with a characteristic L\.{e}vy tail parameter (not
to be confused with the exponent $\alpha$ used in the constraints) defined by
$Pr$($\xi_{i}\geq xt)=x^{-\alpha_{L}}Pr$($\xi_{i}\geq t),$ in the range $1$
$<\alpha_{L}\leq2.$\cite{Kap89} indicating that only a finite expectation
value (first moment) can be defined.

The following example is relevant for non-Debye relaxations. If we derive the
$mepdf$ $\ f(x)$ of the random variable $X$ \ defined in the range
$(0,\infty)$ and if we maximize the S--B entropy subject to the two constraints%

\begin{equation}
<x^{\alpha}>\text{ }=\int_{0}^{\infty}x^{\alpha}f(x)dx\text{ \ \ \ \ \ \ \ \ }%
<\ln x>\text{ }=\int_{0}^{\infty}(\ln x)f(x)dx
\end{equation}
using the Lagrange variational method, it can be shown easily that the
following $mepdf$%

\begin{equation}
f(x)=\frac{\alpha x^{\alpha-1}}{<x^{\alpha}>}\exp[-(\frac{x^{\alpha}%
}{(<x^{\alpha}>)})]
\end{equation}
maximizes the S-B-entropy \cite{Kap89} provided%

\begin{equation}
-\lambda_{2}\ln x=\ln x^{\alpha-1}%
\end{equation}
In that case, if we chose $<x^{\alpha}>=1$ as the natural scale of the problem
the exponent $\alpha$ is related to $\ln x$ as
\begin{equation}
<\ln x>=-\gamma/\alpha
\end{equation}
where $\gamma=\Gamma^{\prime}(1)=0.577215...$ \ is the Euler constant.

The Weibull distribution function is given by%

\[
F(x)=\int_{0}^{\infty}f(x^{\prime})dx^{\prime}=1-\exp(-x^{\alpha}/<x^{\alpha
}>)
\]
As it is well known the Weibull distribution which is used in many physical
problems exhibits a power law behavior $F(x)=x^{\alpha}$ for $x\rightarrow0.$
For $\alpha<1,$ The Williams-Watts formula (stretched exponential) used for
decades to fit the non-Debye relaxation function in the time
domain\cite{Ram87}.

\section{Maximization with Tsallis entropy}

A generalization of the S-B entropy is appropriate when the phenomena
(physical, biological, economical...) are described by distributions with a
L\.{e}vy characteristic parameter $\alpha_{L}<1.$ Here we will start from the
Tsallis non-extensive entropy%

\begin{equation}
S_{q}=-\int_{a}^{b}f_{q}^{q}(x)\ln_{q}f_{q}(x)dx
\end{equation}
subject to the conditions%

\begin{align}
\int_{a}^{b}f_{q}(x)dx  &  =1,\int_{a}^{b}g_{q},_{i}(x)\tilde{f}%
_{q}(x)dx=<g_{q,i}(x)>_{q},\text{ \ }i=1,2,.\ .\\
\tilde{f}_{q}(x)  &  =\frac{f_{q}^{q}(x)}{\int_{a}^{b}f_{q}^{q}(x)dx}\text{
\ \ }f_{q}(x)=\frac{\tilde{f}_{q}^{1/q}(x)}{\int_{a}^{b}\tilde{f}_{q}%
^{1/q}(x)dx}%
\end{align}
The function $\tilde{f}_{q}(x)$ is the so-called \textquotedblright
escort\textquotedblright\ probability \cite{Tsa98}\cite{Tsa01} . Both limits
of integration $a$ and $b$ may be finite or infinite.\ To generalize what has
been done with the standard S-B. entropy we can consider the constraints (3)
using the generalized $q$-exponential and $q$-logarithm functions ($q$-constraints).%

\begin{equation}
\ln_{q}x\equiv\frac{x^{1-q}-1}{1-q}\text{ \ \ \ \ \ }\exp_{q}(x)\equiv
\lbrack1+(1-q)x]^{\frac{1}{1-q}}\text{ \ \ \ with \ \ }\ln_{q}(\exp_{q}(x))=x
\end{equation}
The $mepdf$ depends on the choice of $a$ and $b$ and the functions whose
expectation values are prescribed.

One constructs the Lagrangian
\[
L_{q}=-\int_{0}^{\infty}f_{q}^{q}(x)\ln f_{q}(x)dx-
\]

\begin{equation}
-\lambda_{0}(\int_{0}^{\infty}f_{q}(x)dx-1)-\sum_{i}\lambda_{q,i\ }(\int
_{0}^{\infty}g_{q,i}(x)_{i}\tilde{f}_{q}(x)dx-<g_{qi}(x)>_{q})\
\end{equation}
$\ $

and differentiating with respect to $f_{q}(x)$ one finds%

\begin{equation}
f_{q}(x)=C_{q}\exp_{q}[-\sum_{i}\lambda_{q,i}g_{q,i}(x)]
\end{equation}

\begin{equation}
\tilde{f}_{q}(x)=\tilde{C}_{q}f_{q}(x)^{q}\
\end{equation}

The factors $C_{q}$ are normalization constants and the Lagrange parameters
$\lambda_{q,i}$ are determined by using the $q$-constraints (11). When this
cannot be achieved simply, the parameters $\lambda_{i}$ are parameters defined
by these constraints.

Here we will derive the $mepdf$ $f_{q}(x)$ of the random variable $X$
generalizing \ the two constraints (6) leading to the Weibull distribution%

\begin{equation}
<x^{\alpha}>_{q}=\bar{g}_{1}%
\end{equation}%
\[
<\log_{q}x>_{q}=\bar{g}_{2}%
\]
Using the properties of ln$_{q}(x)$and $\exp_{q}(x)$, we obtain easily the
generalization of the Weibull distribution as a generalized Pareto law
(Burr$XII$ : $B_{b,c}(x)=1-(1+x^{b})^{-c}$, $x>0\ $\cite{Joh70}\cite{Rod77}%

\begin{equation}
f(x)=\frac{\alpha x^{\alpha-1}}{<x^{\alpha}>_{q}}[1+(\frac{q-1}{2-q}%
)\frac{x^{\alpha}}{<x^{\alpha}>_{q}}]^{-\frac{1}{q-1}}%
\end{equation}

\begin{equation}
F(x)=1-[1+(\frac{q-1}{2-q})\frac{x^{\alpha}}{<x^{\alpha}>_{q}}]^{-\frac
{2-q}{q-1}}%
\end{equation}
provided
\begin{equation}
-\lambda_{2}\ln_{q}x=(1+(1-q)\lambda_{1}x^{\alpha})\ln_{q}x^{\alpha-1}%
\end{equation}
which tends to (8) for $q\rightarrow1.$ If $\ $the natural scale $<x^{\alpha
}>_{q}=1,$ one can obtain for that case the following result:%

\begin{equation}
<\ln_{q}x^{\alpha}>_{q}=\frac{(1-q)\Gamma(1/(q-1))+(\frac{q-1}{2-q})^{q}%
\Gamma(3-q)\Gamma(\frac{2-q}{q-1}+q)}{(1-q)(2-q)\Gamma(\frac{2-q}{q-1})}%
\end{equation}
If $q\rightarrow1,$ $<\ln_{q}x^{\alpha}>_{q}\rightarrow-\gamma$ which is the
result (9) obtained for the Weibull distribution. For $\frac{2\alpha+1}%
{\alpha+1}<q<2$ this distribution belongs to the domain of attraction of the
one-sided L\.{e}vy-stable law since we have%

\begin{equation}
\ \lim_{t\rightarrow\infty}\frac{1-F_{q}(xt)}{1-F_{q}(t)}=x^{-\alpha_{L}}%
\end{equation}
with the heavy tail index%

\begin{equation}
\alpha_{L}=\alpha\frac{2-q}{q-1}<1
\end{equation}
In the limit $q\rightarrow1,$ we recover the stretched exponential
Williams-Watts formula.

\section{Universality\ in\ non-Debye\ relaxation}

This result can be used to represent non-Debye relaxation if we identify the
random variable $X$ with the macroscopic waiting time $\tilde{\Theta}$ as
defined by Weron $et.al.$\cite{Wer97}\cite{Wer99}\cite{Jon03}.\ We have :%

\begin{equation}
F_{\Theta}(t)=\Pr(\tilde{\theta}<t)=\int_{0}^{t}f(s)ds=1-[1+(\frac{q-1}%
{2-q})\frac{t^{\alpha}}{<\tilde{\theta}^{\alpha}>_{q}}]^{-\frac{2-q}{q-1}}%
\end{equation}

The relaxation function $\phi$($t)$ can be written as the survival probability
of the non equilibrium initial state of the relaxing system.\ Its value is
determined by the probability that the system as a whole will not make
transition out of its original state until time $t$ :%

\begin{equation}
\phi_{\alpha,q}(t)=\Pr(\tilde{\theta}\geq t)=[1+(\frac{q-1}{2-q}%
)\frac{t^{\alpha}}{<\tilde{\theta}^{\alpha}>_{q}}]^{-\frac{2-q}{q-1}}%
\end{equation}

This expression has the form of one of the universal relaxation functions
proposed in \cite{Wer97}\cite{Wer99} as result of an elaborate study of the
stochastic mechanisms underlying relaxation dynamics in non Debye complex
systems.
\begin{equation}
\phi_{W}(t)=[1+k\ (At)^{\alpha}]^{-1/k}%
\end{equation}
\ with $0<\alpha<1$ and $k\geq\alpha.$

The equivalence\ of the two formulas allows us to relate the parameter $k$
with the non-extensive parameter $q:$%
\begin{equation}
k=\frac{q-1}{2-q}%
\end{equation}
This relation was already derived in \cite{Bro03}. In the limit $q\rightarrow
1,$we recover the stretched exponential Williams-Watts formula used to fit the
time-domain relaxation data \cite{Mon84} \cite{Ram87}.

The $q$-dependent scale parameter $A$ which is related to the the peak
frequency $\omega_{p}$ of the response function in the frequency domain is
materials and $q-$dependent and is now simply related to the escort
probability average
\begin{equation}
A_{q}=<\tilde{\theta}^{\alpha}>_{q}^{-1/\alpha}%
\end{equation}
and becomes the natural finite time scale in a physical system (for example
the dipolar relaxation) which \textquotedblright choose\textquotedblright%
\ \cite{Wer97} the regime where the usual average of the waiting time is
infinite i.e. $\alpha/k\leq1.$

We can write the response function%

\begin{equation}
f(t)=-\frac{d\phi}{dt}=\alpha\ t^{-1}(tA)^{\alpha}[1+\ k(At)^{\alpha
}]^{-\ 1-\frac{1}{k}}%
\end{equation}
which gives the two power-law asymptotic behavior%

\begin{equation}
f_{\alpha}(t,k_{W})=\left\{
\begin{array}
[c]{c}%
\alpha A(At)^{\alpha-1}\text{ \ for }At<<1\\
\\
\alpha Ak^{-1-1/k}(At)^{-\alpha/k-1}\text{ for }At>>1
\end{array}
\right.
\end{equation}
and in the frequency range, where%

\begin{equation}
\chi(\omega)=\int_{0}^{\infty}\exp[-i\omega t]f(t)dt=\chi^{\prime}%
(\omega)+i\chi^{^{\prime\prime}}(\omega)
\end{equation}
obeys the Jonscher universal laws \cite{Jon77}\cite{Jon96}
\begin{equation}
\lim_{\omega\rightarrow\infty}\frac{\chi^{^{\prime\prime}}(\omega)}%
{\chi^{\prime}(\omega)}=\cot(n\pi/2)\text{ \ \ \ \ \ \ \ \ \ \ }\lim
_{\omega\rightarrow0}\frac{\chi^{^{\prime\prime}}(\omega)}{\chi^{\prime
}(0)-\chi^{\prime}(\omega)}=\tan(m\frac{\pi}{2})
\end{equation}
with $n=1-\alpha,$ \ $m=\alpha/k$ and $k=\frac{q-1}{2-q}.$

The $h$-moment of the Burr$XII$ distribution diverges if $m=\alpha/k<h.$
Therefore the expectation value of the random variable $\tilde{\Theta}$
diverges if $m<1$ or $q>\frac{2\alpha+1}{1+\alpha}$.

In the case of dipolar systems, most of the empirical data for the power-law
exponents $n$ and $m$ are in the range [0,1] and can be accounted for using
the heavy-tailed Burr$XII$ waiting-time distribution with $\alpha/k\leq1$ for
which $<\tilde{\theta}>=\infty.$

This presentation which avoids the usual weighted average of a Debye response
function over a relaxation time or relaxation rate distribution confirms the
relation obtained between the empirical exponents $n$ and $m$ and the
\textquotedblright fractal\textquotedblright\ exponent $\alpha$ and the
nonextensive exponent $q$ derived by two of us \cite{Bro03} using more
phenomenological arguments.\ In that model the cluster volume distribution was
obtained from the $\max S_{q}$ principle and a \textquotedblright
fractal\textquotedblright\ relation between the cluster volume and its
relaxation time was assumed.

The Burr$XII$ distribution can be obtained as a smooth mixture of Weibull
distributions compounded with respect to a random scale parameter distributed
with the gamma distribution \cite{Rod77}. Using this fact we can rewrite (25) as%

\begin{equation}
\phi_{\alpha,q}(t)=\int_{0}^{\infty}\exp(-\lambda^{\alpha}t^{\alpha}%
)d\Gamma_{1/k,k}(\lambda^{\alpha}/A^{\alpha})
\end{equation}
where
\begin{equation}
\Gamma_{1/k,k}(\lambda^{\alpha}/A^{\alpha})=\frac{1}{\Gamma(1/k)}\int
_{0}^{\lambda^{\alpha}/A^{\alpha}}(\frac{x}{k})^{1/k-1}e^{-x/k}d(\frac{x}{k})
\end{equation}
is the generalized gamma distribution with scale parameter $k$ and form factor
$\sigma=$ $1/k.$ \cite{Wer97}. It is however to be noticed that this procedure
leads to a relaxation function with a $A$ factor defined as a usual ($q=1)$
expectation value.\ By contrast the $\max S_{q}$\ principle allows to define
$A$ as a natural scale in a situation where the expectation value cannot be defined.

\section{Stochastic theory and Tsallis entropy}

The discussion of the equivalence of the two results obtained for the
\textquotedblright universal\textquotedblright\ relaxation function (25)
and(26) from the $\max S_{q}$ method and the Weron et al. stochastic theory
\cite{Wer97}\cite{Wer99} can lead to a better understanding of the stochastic
properties underlying the Tsallis entropy concept as well as its possible
generalization for more complex spatio-temporal dynamical systems.

The starting point is the observation of the fact that the relaxation function
for the whole system $i.e$ its survival probability is just the probability of
its first passage%

\begin{equation}
\phi(t)=\Pr(\tilde{\theta}\geq t)=\Pr(\min(\theta_{1N}......\theta_{\nu_{N}%
},_{N})
\end{equation}
where%

\begin{equation}
\text{\ \ \ \ \ \ \ \ \ \ }\Pr(\theta_{iN}\geq t)=<e^{-\beta_{iN}%
t}>\text{\ \ \ \ \ }%
\end{equation}

\begin{equation}
\phi(t)=\Pi_{i}^{\upsilon_{N},N}\Pr(\theta_{iN}\geq t)=\Pi_{i}^{\upsilon
_{N},N}\Pr(e^{-\beta_{iN}t}\geq t)=<\exp(-\tilde{\beta}_{N}t)>
\end{equation}
with
\begin{equation}
\tilde{\beta}_{N}=\frac{\sum\limits_{i=1}^{\nu_{N}}\beta_{i,N}}{a_{N}}%
\end{equation}
The quantity $\tilde{\beta}$ is a random relaxation rate corresponding to the
whole (macroscopic) system. The random relaxation rates $\beta_{i,N}$ and
random waiting times $\theta_{i,N}$ are related to the relaxations of the
local individual entities.\ $N$ \ is\ the number of total entities (atoms,
molecules, dynamical clusters) able to relax, $\nu_{N}$ is the number of
relaxing entities participating effectively in the relaxing process . It is
important to notice that $\nu_{N}$ can be a fixed number or a random quantity.

The distribution of (38) can be approximated by the weak limit%

\begin{equation}
\tilde{\beta}=\lim_{N\rightarrow\infty}(a_{N})^{-1}\tilde{\beta}_{N}%
\end{equation}
where $a_{N}$ is the sequence of relevant normalizing constants. On the basis
of limit theorems of probability theory, it is possible to define the
distribution of the limit (39) even if the distribution of the individual
relaxation rates $\beta_{i,N}$ and the number of $\nu_{N}$ of entities
participating effectively\ in the relaxation process are known to a relatively
limited extend only.

The macroscopic relaxing properties are clearly related to the probability
distribution of the summand $\tilde{\beta}$ (39) \ . In the case of a
deterministic fixed number of relaxing entities, Weron $etal$. \cite{Wer97}
\cite{Wer99}have demonstrated, using the extremal value theorem, that if the
$\tilde{\theta}$ are distributed with a stretched exponential distribution,
the only possible distribution for $\tilde{\beta}$\ is a one-sided stable
L\.{e}vy distribution with index $0<\alpha<1$ \ \ and we have:%

\begin{equation}
\Pr(\tilde{\theta}\geq t)=\int_{0}^{\infty}e^{-bt}dL_{\tilde{\beta}}(b)\text{
}=e^{-(At)^{\alpha}}%
\end{equation}
Using an heuristic argument one can understand physically that if the smallest
local waiting time determines the macroscopic relaxation, this should
correspond to the largest relaxation rate. It is not therefore surprising that
the distribution dominated by the highest terms in (38) is for N$\rightarrow
\infty$ the stable one-sided L\.{e}vy distributions which has the property to
be dominated asymptotically by the highest term \cite{Fel66} \cite{Uch99}.

If the number $\nu_{N}$ is a random variable, the scale parameter is a random
variable and the Burr\textit{XII} distribution is obtained if the scale
distribution is the generalized gamma distribution (34). It can be easily
shown by randomizing the scale parameter in (33) that%

\begin{equation}
\Pr(\tilde{\theta}\geq t)=\int_{0}^{\infty}e^{-bt}dF_{\tilde{\beta}%
}(b)=1-B_{\alpha,1/k}^{XII}(k^{1/\alpha}At)=[1+k\ (At)^{\alpha}]^{-1/k}%
\end{equation}
where $F_{\tilde{\beta}}(b)$ is the Mittag-Lefller distribution. In that case
$\tilde{\beta}\ $is a $\upsilon-$stable random variable and stochastic theory
arguments show that this result is obtained if $\nu_{N}$ is distributed with a
generalized negative binomial distribution \cite{Wer97}.%

\begin{equation}
\Pr(\upsilon_{N}=n)=\frac{\Gamma(1/(k+n-1))}{\Gamma(1/k)(n-1)!}(\frac{1}%
{N})^{1/k}(1-\frac{1}{N})^{n-1}{\huge \ \ \ \ {\normalsize n=1,2,...}\ }%
\end{equation}

\section{Conclusions}

In the case of non-Debye relaxation, the short-range interactions, geometric
and dynamic correlations can be accounted for by maximizing the S-B entropy
with adequate constraints. The small clusters and short-time relaxation can be
described by a stretched exponential relaxation function.\ The exponent
$\alpha\leq1$ introduced in the constraints is a measure of the
\textquotedblright non-idealness\textquotedblright\ of the relaxation
processes at this scale\ (fluctuations of the size and intra-cluster
interactions).The system is extensive in the fractal time domain $t^{\alpha}$
and the distribution of the relaxation rates($\beta$) is heavy-tailed with a
Levy tail exponent equal to $\alpha.$

Larger clusters relaxation and long time power law relaxation can be obtained
by the $maxS_{q}$\ principle and depends on the exponent $\alpha/k$ with
$k=\frac{q-1}{2-q}$.\ The $q$ non- extensivity parameter and the heavy tail
properties of the waiting-time $(\theta)$ distribution are directly related to
the random number of relaxing entities at the meso-or microscopic level which
gives rise to fluctuations of the usual extensive (i.e.corresponding to
exponential distributions) time scale. In that sense the Tsallis entropy can
be viewed as a technique to define a finite macroscopic scale when the number
of entities participating to the spatio-temporal dynamical process is
statistically fluctuating due to natural geometric and dynamical constraints.
That scale can be defined as a $q$-expectation value using the concept of
escort probability.\textit{ }This situation is typical of "frustrated" systems
like glasses, polymers and porous materials.\textit{\ }

Although our discussion has been limited to dipolar relaxation, our
conclusions are relevant to a great number of non-exponential decay phenomena
observed in nature \cite{Wil00}\cite{Bec01}.


\begin{thebibliography}{99}                                                                                               %


\bibitem {Kap89}J.N Kapur, \textit{Maximum Entropy Models in Science and
Engineering} , Wiley and Sons, New York (1989).

\bibitem {Ram87}T.V.Ramakrishnan and M.Raj Lakshmi , \textit{Non-Debye
Relaxation in Condensed Matter }(World Scientific Singapore) (1987).

\bibitem {Tsa98}C.Tsallis, R.S.Mendes and A.R.Plastino , \textit{Physica}
\textbf{A261, }534 (1998).

\bibitem {Tsa01}C.Tsallis in \textit{Non extensive statistical mechanics and
thermodynamics} Ed.S.Abe and Y.Okamoto, Springer Ed., Berlin (2001) and
references herein.

\bibitem {Joh70}N.L.Johnson and S.Kotz , \textit{Distributions in Statistics:
Continuous Univariate Distributions 1,2}, Wiley, New York, (1970).

\bibitem {Rod77}R.N.Rodriguez, \textit{Biometrika} \textbf{64,} 129 (1977).

\bibitem {Wer97}K.Weron and M.Kotulski , \textit{J.Stat.Phys.} \textbf{88,} 1241(1997).

\bibitem {Wer99}A.Jurlewicz and K.Weron , \textit{Cell.Mol.Biol.Lett.}%
\textbf{4}, 56 (1999)\textit{.}

\bibitem {Jon03}A.K.Jonscher, A.Judrlewicz an K.Weron,\textit{ Contemporary
Physics} \textbf{44}, 329 (2003).

\bibitem {Bro03}F.Brouers and O.Sotolongo-Costa , \textit{Europhys
Lett.}\textbf{62}, 808 (2003)\textit{. }

\bibitem {Mon84}E.W.Montroll and J.T.Bendler, \textit{J.Stat.Phys.}
\textbf{34,} 129 (1984).

\bibitem {Jon77}A.K.Jonscher, \textit{Nature } \textbf{267,} 673 (1977).

\bibitem {Jon96}A.KJonscher, \textit{Universal Relaxation laws, }Chelsea
Dielectric Press, London (1996).

\bibitem {Uch99}V.V.Uchaikin and V.M.Zolotarev, \textit{Chance and Stability}
VSP (Utrecht) (1999).

\bibitem {Fel66}Feller J., \textit{An introduction to Probability Theory and
Its Applications,} 2, Wiley, New York (1966).

\bibitem {Wil00}G.Wilk and Wlodarczyk, \textit{Phys.Rev.Letters} \textbf{84},
2770 (2000).

\bibitem {Bec01}C.Beck\textit{Phys.Rev.Letters} \textbf{87}, 18060 (2001).
\end{thebibliography}
\end{document}